\documentclass[12pt]{iopart}
 
\expandafter\let\csname equation*\endcsname\relax
\expandafter\let\csname endequation*\endcsname\relax
\usepackage{suffix}
\usepackage{mathtools}

\DeclarePairedDelimiterX\MeijerM[3]{\lparen}{\rparen}%
{\begin{smallmatrix}#1 \\ #2\end{smallmatrix}\delimsize\vert\,#3}

\newcommand\MeijerG[8][]{%
  G^{\,#2,#3}_{#4,#5}\MeijerM[#1]{#6}{#7}{#8}}

\WithSuffix\newcommand\MeijerG*[7]{%
  G^{\,#1,#2}_{#3,#4}\MeijerM*{#5}{#6}{#7}}

\renewcommand{\vec}[1]{\mathbf{#1}}
\newcommand{\stirling}[2]{\left\{\begin{matrix}#1\\#2\end{matrix}\right\}}

\begin{document}

\title[The model of a level crossing with a Coulomb band]{The model of a level crossing with a Coulomb band: exact probabilities of nonadiabatic transitions}

\author{J Lin$^{1,2}$ and N A Sinitsyn$^2$}
\address{$^1$Department of Mathematics, Princeton University, Princeton, NJ 08544, USA}
\address{$^2$Theoretical Division,
Los Alamos National Laboratory,
Los Alamos, NM 87545, USA}
\ead{jeffminl@princeton.edu, nsinitsyn@lanl.gov}

\begin{abstract}
We derive an exact solution of an explicitly time-dependent multichannel model of quantum mechanical nonadiabatic transitions. Our model corresponds to the case of a single linear diabatic energy level interacting with a band of an arbitrary $N$ states, for which the diabatic energies decay with time according to the Coulomb law. We show that the time-dependent Schr\"odingier equation for this system can be solved in terms of Meijer functions whose asymptotics at a large time can be compactly written in terms of elementary functions that depend on the roots of an $N$th order characteristic polynomial. Our model can be considered a generalization of the Demkov-Osherov model. In comparison to the latter, our model allows one to explore the role of curvature of the band levels and diabatic avoided crossings.
\end{abstract}

\maketitle

\section{Introduction}
Models that describe multi-channel nonadiabatic processes in explicitly time-dependent quantum systems are important for the theory of control and characterization of numerous mesoscopic and atomic systems \cite{book}. Several exact results for explicitly driven two state systems, such as the Landau-Zener-Stuekkelberg-Majorana formula \cite{LZ}, have become widely used in practice. However, the physics of nonadiabatic transitions in {\it multichannel} explicitly driven systems is a much more complex topic. Exactly solvable models can provide the needed intuition about this subject. They also can be used as a starting point for various approximation schemes and testing numerical algorithms and analytical methods.

A relatively large class of exactly solvable Landau-Zener-like models with multichannel interactions has been identified and explored in \cite{be,  mlz-1, mlz-2,  do, reducible, bow-tie,ostrovsky, sinitsyn-13prl1}. These models correspond to quantum mechanical evolution of systems with matrix Hamiltonian operators of the type
\begin{equation}
\hat{H}(t) = \hat{A} +\hat{B}t + \frac{\hat{C}}{t},
\label{ham2}
\end{equation}
where $\hat{A}$, $\hat{B}$ and $\hat{C}$ are constant $N\times N$ matrices. One has to find the scattering $N\times N$ matrix $\hat{S}$, in which the element $S_{nn'}$ is the amplitude of the diabatic state $n'$ at $t \rightarrow +\infty$, given that at $t \rightarrow -\infty$ (or at $t=0$) the system was at the state $n$.  The related matrix $\hat{P}$, with $P_{nn'}=|S_{nn'}|^2$, is called the matrix of transition probabilities.

In this article, we present a new exactly solvable system of the type \eref{ham2} and determine its matrix of transition probabilities. 
Schr\"odinger's equation for the amplitudes, called $b_0$ and $a_i$, $i=1\ldots N$, of this model is given by
\begin{align}
i\frac{db_0}{d\tau}&=\beta\tau b_0 +\sum_{j=1}^N g_ja_j\label{eq:sch1}\\
i\frac{da_j}{d\tau}&=\frac{k_j}{\tau}a_j+g_jb_0,\quad 1\leq j\leq N,\label{eq:sch2}
\end{align}
with constant parameters $\beta$, $k_i$ and $g_i$. We will call $b_0(t)$ the amplitude of the ``0th level,'' $a_i(t)$ the amplitude of the ``$i$th level,'' etc.; parameters $g_i$  are called the coupling constants, and $\beta$ is called the slope of the $0$th diabatic energy level. 

\begin{figure}
\begin{indented}
\item[]\includegraphics[scale=.15]{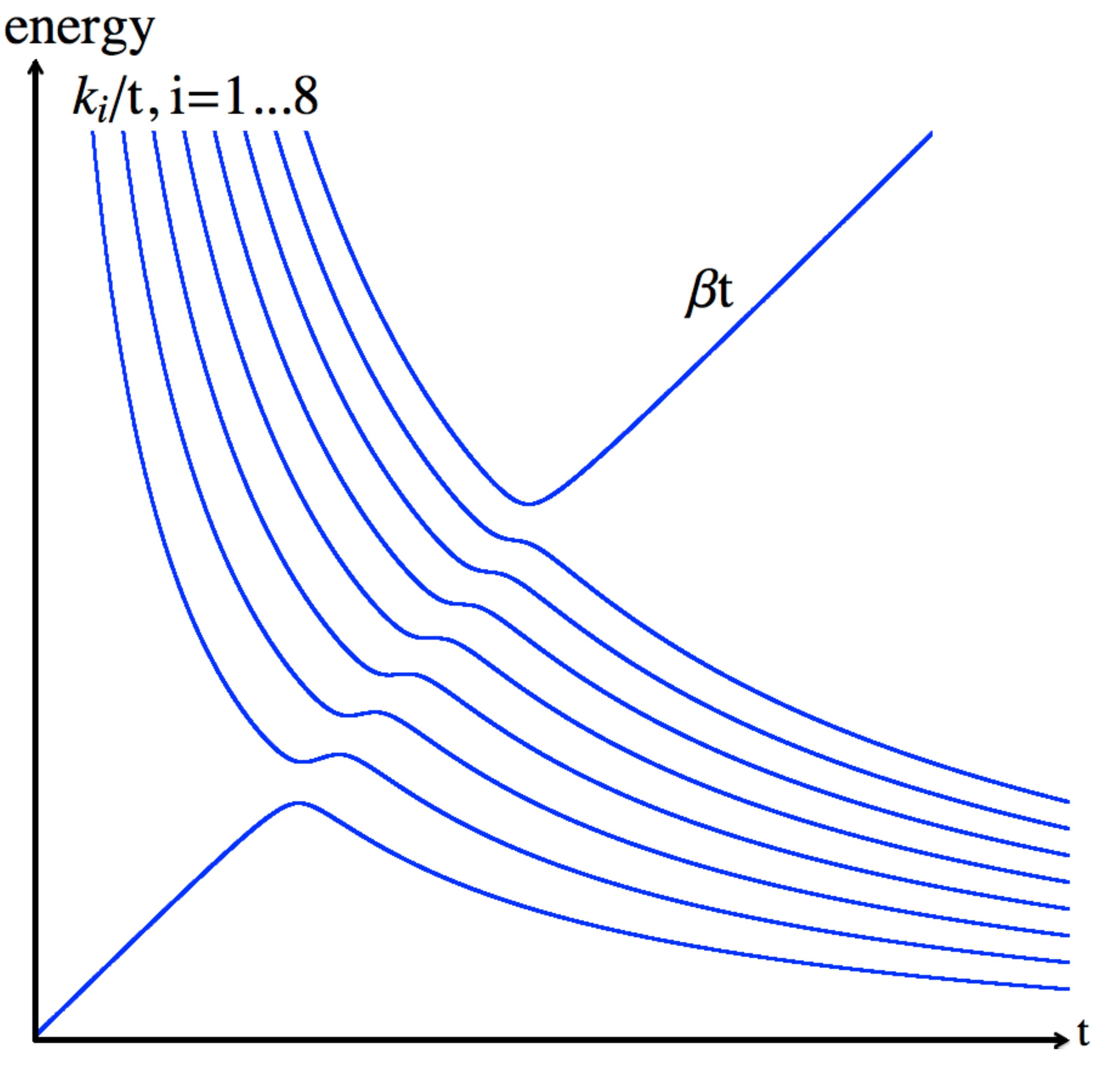}
\end{indented}
\caption{\label{rays} Time-dependence of the eigenspectrum (adiabatic energies) of the Hamiltonian with elements $H_{00}=\beta t$, $H_{nn} = k_n/t$, $H_{n0}=H_{0n}=g_n$, and zero otherwise ($n=1,\ldots, N$; $N=8$). For any $n$, all $k_n>0$.}
\end{figure}

This model corresponds to the case of a single linear diabatic energy level crossing a band of states whose diabatic energies (diagonal elements of the Hamiltonian matrix) decay as $\sim k_i/t$ with time (\fref{rays}). Structurally, the model (\ref{eq:sch1}, \ref{eq:sch2}) is very similar to the celebrated Demkov-Osherov model \cite{do}. The latter corresponds to the case of a single level crossing a band of parallel states. In fact, one can show that the Demkov-Osherov solution is recovered from our model in a specific limit: $k_i \gg |k_i-k_j| \sim |g_i|^2/\beta$, $i,j=1\ldots N$.  When this condition is not satisfied, the band has a nonzero curvature near the avoided level crossings. Hence, our solution makes a unique insight into the physics of nonadiabatic transitions in the case when a band has a non-zero curvature, as shown in \fref{rays}.
Another interesting situation corresponds to the case when all $k_i$ are negative. Our model then corresponds to the system with nonadiabatic transitions between a level that passes in the vicinity of a band but does not go through avoided level crossings (figure~\ref{diab-av}).

We will discuss that not all transition probabilities among diabatic states are well defined in this system because states of the band become asymptotically degenerate at $t\rightarrow \infty$, so that transitions among them never saturate.
 Therefore, our main focus will be on the probability of transition to the 0th level. In particular, we will present a simple formula (Eq.~\ref{eq:p00}) for the probability to remain in this level after all interactions if this level was initially  populated.  

\section{Solution of the model}

\subsection{Transforming \Eref{eq:sch1} into the Meijer equation}

We may assume without loss of generality that the $k_j$ are ordered with $k_1<\cdots<k_N$.
The change of variables $t=\tau^2/2$, $a_j=\tau b_j$ gives:
\begin{align}
i\frac{db_0}{dt}&=\beta b_0+\sum_{j=1}^Ng_jb_j\label{eq:elim1}\\
2it\frac{db_j}{dt}&=(k_j-i)b_j+g_jb_0,\quad 1\leq j\leq N.\label{eq:elim2}
\end{align}
It follows from \eref{eq:elim2} that
\begin{equation}
\left[t\frac{d}{dt}+\left(\frac{1}{2}+i\frac{k_j}{2}\right)\right]b_j=\frac{g_j}{2i}b_0.\label{eq:elim3}
\end{equation}
Then applying \eref{eq:elim3} to \eref{eq:elim1}, we have
\begin{equation}
i\prod_{j=1}^N\left[t\frac{d}{dt}+\left(\frac{1}{2}+i\frac{k_j}{2}\right)\right]\frac{d}{dt}b_0=\beta\prod_{j=1}^N\left[t\frac{d}{dt}+\left(\frac{1}{2}+i\frac{k_j}{2}\right)\right]b_0+\sum_{j=1}^N\frac{g_j^2}{2i}\prod_{\substack{m=1\\ m\neq j}}^N\left[t\frac{d}{dt}+\left(\frac{1}{2}+i\frac{k_m}{2}\right)\right]b_0.\label{eq:elim4}
\end{equation}
Define the following polynomial in $x$:
\begin{equation}
f(x):=\beta\prod_{j=1}^N\left[x-\left(\frac{1}{2}-i\frac{k_j}{2}\right)\right]+\sum_{j=1}^N\frac{g_j^2}{2i}\prod_{\substack{m=1\\ m\neq j}}^N\left[x-\left(\frac{1}{2}-i\frac{k_m}{2}\right)\right].\label{eq:poly1}
\end{equation}
Letting $x=1/2+iy$, we define the following function of $y$:
\begin{equation}
g(y):=\frac{(-i)^N}{\beta}f(x)=\prod_{j=1}^N\left[y+\frac{k_j}{2}\right]-\sum_{j=1}^N\frac{g_j^2}{2\beta}\prod_{\substack{m=1\\ m\neq j}}^N\left[y+\frac{k_m}{2}\right].\label{eq:poly2}
\end{equation}
Because the sign of the ordered set $\{g(-\infty),g(-k_N/2),\ldots,g(-k_1/2),g(\infty)\}$ alternates exactly $N$ times, $g$ has exactly $N$ real roots. Let these roots be $l_1,\ldots,l_N$, and let $\xi_j=1/2+il_j$ be the corresponding roots of $f(x)$. We note that
\begin{equation}
\sum_{j=1}^N l_j=\sum_{j=1}^N\left(\frac{g_j^2}{2\beta}-\frac{k_j}{2}\right).\label{eq:poly3}
\end{equation}

\begin{figure}
\begin{indented}
\item[]\includegraphics[scale=.25]{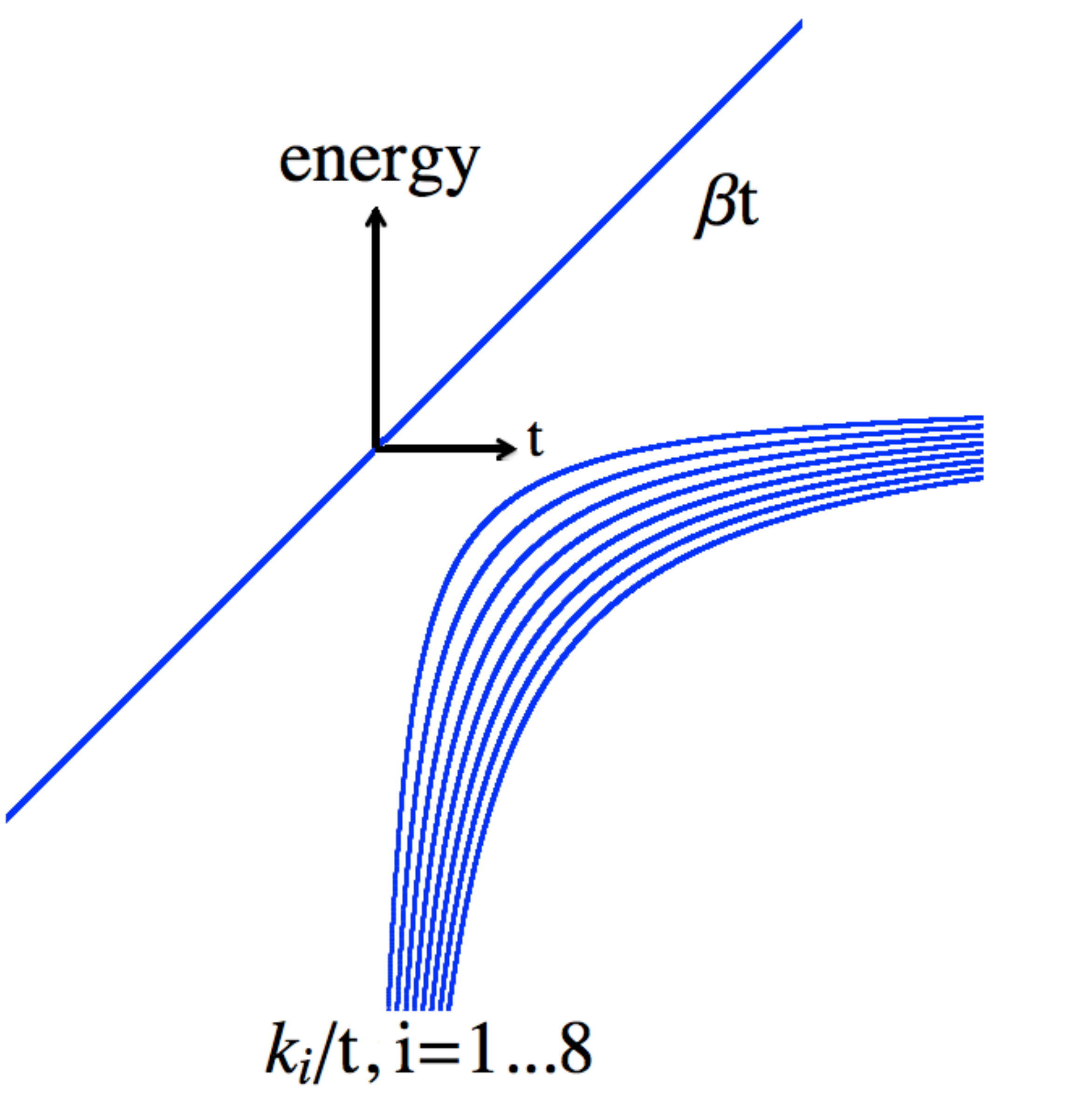}
\end{indented}
\caption{\label{rays5} Time-dependence of the eigenspectrum (adiabatic energies) of the Hamiltonian with elements $H_{00}=\beta t$, $H_{nn} = k_n/t$, $H_{n0}=H_{0n}=g_n$, and zero otherwise ($n=1,\ldots, N$; $N=8$). For any $n$, all $k_n<0$.
 }
 \label{diab-av}
\end{figure}
With these definitions in place, \eref{eq:elim4} becomes
\begin{equation}
i\prod_{j=1}^N\left[t\frac{d}{dt}+\left(\frac{1}{2}+i\frac{k_j}{2}\right)\right]\frac{d}{dt}b_0=f\left(t\frac{d}{dt}+1\right).\label{eq:elim5}
\end{equation}
Because the operator
\begin{equation}
t\frac{d}{dt}+1\label{eq:poly4}
\end{equation}
commutes with itself and multiplication by a constant, \eref{eq:elim5} becomes
\begin{equation}
\left(i\prod_{j=1}^N\left[t\frac{d}{dt}+\left(\frac{1}{2}+i\frac{k_j}{2}\right)\right]\frac{d}{dt}-\beta\prod_{j=1}^N\left[t\frac{d}{dt}+1-\xi_j\right]\right)b_0=0.\label{eq:elim6}
\end{equation}
We use the identity
\begin{equation}
t\frac{d^2}{dt^2}=\frac{d}{dt}\left(t\frac{d}{dt}-1\right)\label{eq:elim7}
\end{equation}
and multiply \eref{eq:elim6} by $it$ to get the desired form of the differential equation for $b_0$:
\begin{equation}
\left(-t\frac{d}{dt}\prod_{j=1}^N\left[t\frac{d}{dt}-\left(\frac{1}{2}-i\frac{k_j}{2}\right)\right]-\beta it\prod_{j=1}^N\left[t\frac{d}{dt}+1-\xi_j\right]\right)b_0=0.\label{eq:b0diffeq}
\end{equation}
\Eref{eq:b0diffeq} is an $(N+1)$th order linear differential equation, which can be found in Section 5.8 of \cite{luke}, equation 1. It has a fundamental set of $N+1$ Meijer's G solutions.

\subsection{Solving for $b_0$}

For notational convenience, define
\begin{equation}
h_j=\frac{1}{2}-i\frac{k_j}{2},\label{eq:note1}
\end{equation}
and the following three sets:
\begin{align}
\vec{\xi_N}&=\xi_1,\ldots,\xi_N\label{eq:note2}\\
\vec{h_N}&=h_1,\ldots,h_N\label{eq:note3}\\
\vec{h_N}^r&=h_1,\ldots,h_{r-1},h_{r+1},\ldots,h_N.\label{eq:note4}
\end{align}
Equation \eref{eq:b0diffeq} has the following fundamental set of $N+1$ Meijer's G solutions:
\begin{equation}
\MeijerG*{1}{N}{N}{N+1}{\vec{\xi_N}}{0,\vec{h_N}}{\beta it},\left\{\MeijerG*{1}{N}{N}{N+1}{\vec{\xi_N}}{h_r,0,\vec{h_N}^r}{\beta it}\right\}_{r=1}^{r=N}.\label{eq:b0fundset}
\end{equation}
Each of these solutions remains bounded as $t\rightarrow 0$, and furthermore, all but the first go to zero as $t\rightarrow 0$ (\ref{calcs}\ref{calc1}).

\subsection{Finding the vector of solutions}

Each of the solutions in \eref{eq:b0fundset} for $b_0$ specifies a unique vector of solutions to the original system of equations, and the general solution to the system (\ref{eq:elim1}, \ref{eq:elim2}) is a linear combination of these vectors. Manipulating the system (\ref{eq:elim1}, \ref{eq:elim2}) gives the following $N$ equations:
\begin{align}
\notag\sum_{j=1}^N g_jb_j &= i\frac{db_0}{dt}-\beta b_0\\\notag
\sum_{j=1}^N g_j(k_j-i)b_j &= 2it\frac{d}{dt}\left[i\frac{db_0}{dt}-\beta b_0\right]-\sum_{j=1}^Ng_j^2b_0\\\notag
\sum_{j=1}^N g_j(k_j-i)^2b_j &= 2it\frac{d}{dt}\left[2it\frac{d}{dt}\left[i\frac{db_0}{dt}-\beta b_0\right]-\sum_{j=1}^Ng_j^2b_0\right]-\sum_{j=1}^Ng_j^2(k_j-i)b_0\\
&\vdots\notag\\
\sum_{j=1}^N g_j(k_j-i)^{N-1}b_j &= 2it\frac{d}{dt}\left[\cdots\left[2it\frac{d}{dt}\left[i\frac{db_0}{dt}-\beta b_0\right]-\sum_{j=1}^Ng_j^2b_0\right]\cdots\right]-\sum_{j=1}^Ng_j^2(k_j-i)^{N-2}b_0.
\end{align}
Multipying through by $(2t)^{1/2}$ gives equations for the $a_j$:
\begin{align}
\notag\sum_{j=1}^N g_ja_j &= (2t)^{1/2}\left(i\frac{db_0}{dt}-\beta b_0\right)\\\notag
\sum_{j=1}^N g_j(k_j-i)a_j &= (2t)^{1/2}\left(2it\frac{d}{dt}\left[i\frac{db_0}{dt}-\beta b_0\right]-\sum_{j=1}^Ng_j^2b_0\right)\\\notag
\sum_{j=1}^N g_j(k_j-i)^2a_j &= (2t)^{1/2}\left(2it\frac{d}{dt}\left[2it\frac{d}{dt}\left[i\frac{db_0}{dt}-\beta b_0\right]-\sum_{j=1}^Ng_j^2b_0\right]-\sum_{j=1}^Ng_j^2(k_j-i)b_0\right)\\
&\vdots\notag\\
\sum_{j=1}^N g_j(k_j-i)^{N-1}a_j &= (2t)^{1/2}\left(2it\frac{d}{dt}\left[\cdots\left[2it\frac{d}{dt}\left[i\frac{db_0}{dt}-\beta b_0\right]-\sum_{j=1}^Ng_j^2b_0\right]\cdots\right]-\sum_{j=1}^Ng_j^2(k_j-i)^{N-2}b_0\right)\label{eq:init1-4}.
\end{align}
The left-hand side of \eref{eq:init1-4} is an $N\times N$ Vandermonde matrix
\begin{equation}
V_N=\begin{bmatrix}
1&1&\cdots&1\\
k_1-i&k_2-i&\cdots&k_N-i\\
(k_1-i)^2&(k_2-i)^2&\cdots&(k_N-i)^2\\
\vdots&\vdots&\ddots&\vdots\\
(k_1-i)^{N-1}&(k_2-i)^{N-1}&\cdots&(k_N-i)^{N-1}
\end{bmatrix},\label{eq:initvand}
\end{equation}
(whose inverse is well known; for example, a general expression for the inverse can be found in \cite{bonilla}), multiplied with the column vector $[g_1a_1,\ldots,g_Na_N]^T$.

\section{Transition probabilites}
\subsection{Initially populating $b_0$}
Consider the first solution in \eref{eq:b0fundset}:
\begin{equation}
\eqalign{b_0&=\MeijerG*{1}{N}{N}{N+1}{\vec{\xi_N}}{0,\vec{h_N}}{\beta it}\\
&=\prod_{j=1}^N\left(\frac{\Gamma(1-\xi_j)}{\Gamma\left(\frac{1}{2}+i\frac{k_j}{2}\right)}\right){}_NF_N\left(1-\xi_1,\ldots,1-\xi_N;\frac{1}{2}+i\frac{k_1}{2},\ldots,\frac{1}{2}+i\frac{k_N}{2};-\beta it\right)}
\label{eq:b0firstsol}.
\end{equation}
Then $b_0(t\rightarrow 0)<\infty$, and since the $k$th derivative of a hypergeometric function is a multiple of another hypergeometric function (\ref{calcs}\ref{calc2}), we have
\begin{equation}
\left.\frac{d^k}{dt^k}b_0\right|_{t\rightarrow 0}<\infty.
\end{equation}
Then it follows from  \eref{eq:init1-4} that with this choice of $b_0$, $a_1(0)=\cdots=a_N(0)=0$. We now set
\begin{equation}
b_0=\prod_{j=1}^N\left(\frac{\Gamma\left(\frac{1}{2}+i\frac{k_j}{2}\right)}{\Gamma(1-\xi_j)}\right)\MeijerG*{1}{N}{N}{N+1}{\vec{\xi_N}}{0,\vec{h_N}}{\beta it}\label{eq:b0initpop},
\end{equation}
in order that $b_0$ may be initially populated with probability 1, and using \eref{eq:b0firstsol} and the asymptotics from \eref{asymp1}, we find (\ref{calcs}\ref{calc3}): 
\begin{equation}
P_{00}=|b_0(t\rightarrow\infty)|^2=\prod_{j=1}^N\left(\frac{\exp(-2\pi l_j)+1}{\exp(\pi k_j)+1}\right).\label{eq:p00} 
\end{equation}

With $a_1(0)=\cdots=a_N(0)=0$ and $b_0$ chosen as in \eref{eq:b0initpop}, we may simply solve \eref{eq:elim2} for each $j$, using the antiderivative in \eref{deriv1}, to find
\begin{align}
a_j&=\frac{-iQg_jt^{-ik_j/2}}{\sqrt{2}}\int_0^t x^{(1-h_j)-1}\MeijerG*{1}{N}{N}{N+1}{\vec{\xi_N}}{0,\vec{h_N}}{\beta ix}\ dx\\
&=\frac{-iQg_jt^{-ik_j/2}}{\sqrt{2}}t^{1-h_j}\MeijerG*{1}{N}{N}{N+1}{\vec{\xi_N}}{0,\vec{h_N}^j,h_j-1}{\beta it}\label{eq:b0toajsol},
\end{align}
where
\begin{equation}
Q=\prod_{j=1}^N\left(\frac{\Gamma\left(\frac{1}{2}+i\frac{k_j}{2}\right)}{\Gamma(1-\xi_j)}\right)\label{eq:b0solfactor}.
\end{equation}
The resultant asymptotic is (\ref{calcs}\ref{calc4}):
\begin{equation}
\eqalign{P_{0j}&=|a_j|^2\\
&\sim\frac{|Q|^2g_j^2}{2\beta}\left|\sum_{s=1}^N e^{-\pi l_s/2}(\beta t)^{il_s}\frac{\Gamma(1-\xi_s)}{\Gamma(1+\xi_s-h_j)}\left(\frac{\prod_{r=1,r\neq s}^N\Gamma(\xi_s-\xi_r)}{\prod_{r=1,r\neq j}^N\Gamma(\xi_s-h_r)}\right)\right|^2.}\label{eq:p0j}
\end{equation}
Equation \eref{eq:p0j} shows that, unlike $P_{00}$, the probability $P_{0j}$ does not converge at $t\rightarrow\infty$ in the general case. Physically, this behavior follows from the fact that diabatic energies of levels with $j\ne0$ become asymptotically degenerate at $t\rightarrow \infty$ with a characteristic level splitting behaving as $\sim 1/t$ with time. On the other hand, couplings between those levels, in the leading order of the perturbation expansion, also decay as $\sim 1/t$, so that both diagonal and off-diagonal terms in the Hamiltonian projected on the N state subspace of levels with $j \ne 0$ remain of the same order in magnitude. Hence transitions between such diabatic states never saturate. One can explore the problem of scattering in the adiabatic basis but we will not pursue it here. 

\subsection{Initially populating $a_q$ to find $P_{q0}$ (arbitrary N)}

As mentioned earlier (see \ref{calcs}\ref{calc1}) all but the first solution in \eref{eq:b0fundset} go to 0 when $t\rightarrow 0$; furthermore, by \eref{deriv2}, it follows that
\begin{equation}
\left.t^{k-1/2}\frac{d^k}{dt^k}b_0\right|_{t\rightarrow 0}<\infty,
\end{equation}
so in the limit $t\rightarrow 0$, \eref{eq:init1-4} reduces to
\begin{equation}
\left\{\sum_{j=1}^N g_j(k_j-i)^{m-1}a_j(0) = \left[(2t)^{1/2}i(2i)^{m-1}\left[\left(t\frac{d}{dt}\right)^{m-1}\frac{db_0}{dt}\right]\right]_{t\rightarrow 0}\right\}_{m=1}^{N}.\label{eq:cond1-4}
\end{equation}
We have that
\begin{equation}
\left(t\frac{d}{dt}\right)^m=\sum_{j=1}^m \stirling{m}{j}t^j\frac{d^j}{dt^j},
\end{equation}
where
\begin{equation}
\stirling{m}{j}=S(m,j)\label{eq:gencoeff}
\end{equation}
denotes the Stirling numbers of the second kind, which satisfy the recurrence
\begin{equation}
\stirling{m+1}{j} = j\stirling{m}{j} + \stirling{m}{j-1},
\end{equation}
and are equal to the number of ways to partition $m$ labelled elements into $j$ nonempty unlabelled sets. Then \eref{eq:cond1-4} becomes
\begin{equation}
\left\{\sum_{j=1}^N g_j(k_j-i)^{m-1}a_j(0) = \left[(2t)^{1/2}i(2i)^{m-1}\left[\sum_{j=1}^{m-1} \stirling{m-1}{j}t^j\frac{d^{j+1}}{dt^{j+1}}b_0\right]\right]_{t\rightarrow 0}\right\}_{m=1}^N.\label{eq:cond5-8}
\end{equation}

If we put
\begin{equation}
b_0=t^{ik_r/2}\MeijerG*{1}{N}{N}{N+1}{\vec{\xi_N}}{h_r,0,\vec{h_N}^r}{\beta it},\label{eq:pj0init}
\end{equation}
we have the following result, using \eref{deriv2} and \eref{elem3}:
\begin{align}
\lim_{t\rightarrow 0}t^{j+1/2}\frac{d^{j+1}}{dt^{j+1}}b_0&=\left(\Gamma(h_r-j)\right)^{-1}\prod_{s=1}^N\left(\frac{\Gamma(1+h_r-\xi_s)}{\Gamma(1+h_r-h_s)}\right)\\
&=(h_r-1)_{j}\left(\Gamma(h_r)\right)^{-1}\prod_{s=1}^N\left(\frac{\Gamma(1+h_r-\xi_s)}{\Gamma(1+h_r-h_s)}\right),\label{eq:b0initderiv}
\end{align}
where $(h_r-1)_{j}$ denotes the falling factorial $(h_r-1)\cdots(h_r-j+1)(h_r-j)$ and where $(h_r-1)_0=h_r-1$. Then \eref{eq:cond5-8} becomes
\begin{equation}
\left\{\sum_{j=1}^N g_j(k_j-i)^{m-1}a_j(0) = \gamma_{m,r}\right\}_{m=1}^N,\label{eq:cond9-12}
\end{equation}
where, for notational convenience, we let
\begin{equation}
\gamma_{m,r}=(2)^{m-1/2}(i)^{m}\left(\Gamma(h_r)\right)^{-1}\left[\sum_{j=1}^{m-1} \stirling{m-1}{j}(h_r-1)_{j}\right]\prod_{s=1}^N\left(\frac{\Gamma(1+h_r-\xi_s)}{\Gamma(1+h_r-h_s)}\right).\label{eq:abbrev1}
\end{equation}
The Stirling numbers of the second kind satisfy the relation
\begin{equation}
\sum_{j=1}^{m}\stirling{m}{j}(x)_j = x^{m-1},
\end{equation}
which simplifies \eref{eq:abbrev1} to
\begin{equation}
\gamma_{m,r}=(2)^{m-1/2}(i)^{m}\left(\Gamma(h_r)\right)^{-1}(h_r-1)^{m-1}\prod_{s=1}^N\left(\frac{\Gamma(1+h_r-\xi_s)}{\Gamma(1+h_r-h_s)}\right).\label{eq:abbrev2}
\end{equation}
Using notation from \eref{eq:initvand}, with $b_0$ as in \eref{eq:pj0init},
\begin{equation}
\begin{bmatrix}
g_1a_1(0)\\
\vdots\\
g_Na_N(0)
\end{bmatrix}=V_N^{-1}\begin{bmatrix}
\gamma_{1,r}\\
\vdots\\
\gamma_{N,r}
\end{bmatrix}.\label{eq:geninit}
\end{equation}
Setting $\zeta_r$ (a vector) equal to the right-hand side of \eref{eq:geninit}, we let $\vec{c}_q$ be the $q$th column of
\begin{equation}
\begin{bmatrix}
\vdots&&\vdots\\
\zeta_1&\cdots&\zeta_N\\
\vdots&&\vdots
\end{bmatrix}^{-1}\begin{bmatrix}
g_1&0&\cdots&0\\
0&g_2&\cdots&0\\
\vdots&\vdots&\ddots&\vdots\\
0&0&\cdots&g_N
\end{bmatrix},
\end{equation}
and denote the components of $\vec{c}_q$ by
\begin{equation}
\vec{c}_q=\begin{bmatrix}
c_{1,q}\\
\vdots\\
c_{N,q}
\end{bmatrix}.
\end{equation}
If we put
\begin{equation}
b_0=\sum_{r=1}^N c_{r,q}t^{ik_r/2}\MeijerG*{1}{N}{N}{N+1}{\vec{\xi_N}}{h_r,0,\vec{h_N}^r}{\beta it}\label{eq:bqinitpop},
\end{equation}
this corresponds to the vector of states where $b_0(0)=a_1(0)=\cdots=a_{q-1}(0)=a_{q+1}(0)=\cdots=a_N(0)=0$, and $a_q(0)=1$.

Again applying the asymptotics from \eref{asymp1} provides an asymptotic for $P_{q0}$ (\ref{calcs}\ref{calc5}):
\begin{equation}
P_{q0}=|b_0|^2\sim\exp\left(-\pi\sum_{j=1}^N\frac{g_j^2}{2\beta}\right)\left|\sum_{r=1}^N c_{r,q}t^{ik_r/2}e^{\pi k_r/2}\right|^2.
\end{equation}

Amplitudes for the states $a_j$ are given in \ref{amplitudeqj}.

\section{Discussion}
The solution of our model is expressed through the roots of the polynomial $g$ in \eref{eq:poly2}, which generally cannot be obtained explicitly. To provide better intuition about the transition probabilities in our model, we will explore special situations that allow us to obtain explicit expressions for the transition probabilities.

\subsection{Degenerate band}

Consider the case of all the $k_i$ identical, i.e. $k_i=k$ for all $i=1\ldots N$. This case corresponds to 
\begin{equation}
g(y)=\left(y+\frac{k}{2}\right)^{N-1} \left(y+\frac{k}{2}-\sum_{i=1}^N \frac{g_i^2}{2\beta} \right),
\label{g-deg}
\end{equation}
with simple roots $l_i=-k/2$ for $i=1,\ldots, N-1$ and $l_N=-k/2+\sum_{i=1}^N g_i^2/2\beta$. Substituting this into (\ref{eq:p00}) we find
\begin{equation}
P_{00}=\left(\frac{\exp\left(\pi \left[k-\sum \limits_{i=1}^N\frac{g_i^2}{\beta}\right]\right) +1}{\exp(\pi k)+1}\right).\label{eq:p00-4} 
\end{equation}
Interestingly, this solution shows that, no matter how large $N$ is and how strong coupling constants are, if the band levels are degenerate the survival probability is always non-vanishing, i.e. $P_{00}> 1/\left(e^{\pi k}+1\right)$. Such a behavior can occur only when the curvature of the band is non-vanishing. It was first noticed in the Demkov-Osherov model with a piecewise linearly changing slope of the 0th level \cite{yurovsky}. It is confirmed  now for a  model with fully continuous time-dependence of diabatic energy levels. We also note that, in the case of a degenerate band, a linear transformation can reduce the degenerate states to one coupled to the 0th state and $N-1$ uncoupled, leading to a two-state curve-crossing problem \cite{yurovsky-2}.

\subsection{Independent crossings}

In another limit, consider that $|k_i-k_j| \gg |g_s^2/\beta|$ for any $i,j,s$. This case corresponds to well separated diabatic energies of band levels.
Treating couplings as perturbations, we find
\begin{equation}
l_j \approx -\frac{k_j}{2}+\frac{g_j^2}{2\beta}.
\end{equation}
and 
\begin{equation}
P_{00} \approx \prod \limits_{i=1}^N \left(\frac{\exp\left(\pi \left[k_i-\frac{g_i^2}{\beta} \right]\right) +1}{\exp(\pi k_i)+1}\right), \label{eq:p00-41} 
\end{equation}
i.e. the survival probability is given by the product of probabilities to remain at the $0$th level at each pairwise avoided crossing. Here we note that for any finite $N$ one still has the restriction $P_{00}> 1/\left(\prod_{i=1}^N \left[e^{\pi k_i}+1\right]\right)$; however, by increasing $N$ one can make $P_{00}$ arbitrarily small.

\subsection{The case $N=2$}

When $N=2$, one can obtain an explicit expression for the roots of the polynomial $g(y)$,
\begin{equation}
g(y)=\left(y+\frac{k_1}{2} \right) \left(y+\frac{k_2}{2} \right) -\frac{1}{2\beta} \left( g_1^2\left(y+\frac{k_2}{2} \right) +g_2^2 \left(y+\frac{k_1}{2} \right)  \right),
\label{gy3}
\end{equation}
which are given by
\begin{equation}
l_{1,2}=\frac{g_+ -\beta k_+ \pm \sqrt{ g_+^2+\beta k_{-} (\beta k_{-} -2g_{-})}}{4\beta},
\label{roots3}
\end{equation}
where $k_{\pm} = k_1 \pm k_2$ and  $g_{\pm} =g_1^2 \pm g_2^2$. It is straightforward to check that at $k_{-}\gg g_i^2/\beta$ one has 
$l_{1,2} \approx (g^2_{1,2}/\beta - k_{1,2})/2$, or that at $k_{-}=0$, one has $l_1 = (g_1^2+g_2^2)/\beta$ and $l_2=0$. 

\begin{figure}
\begin{indented}
\item[]\includegraphics[scale=.19]{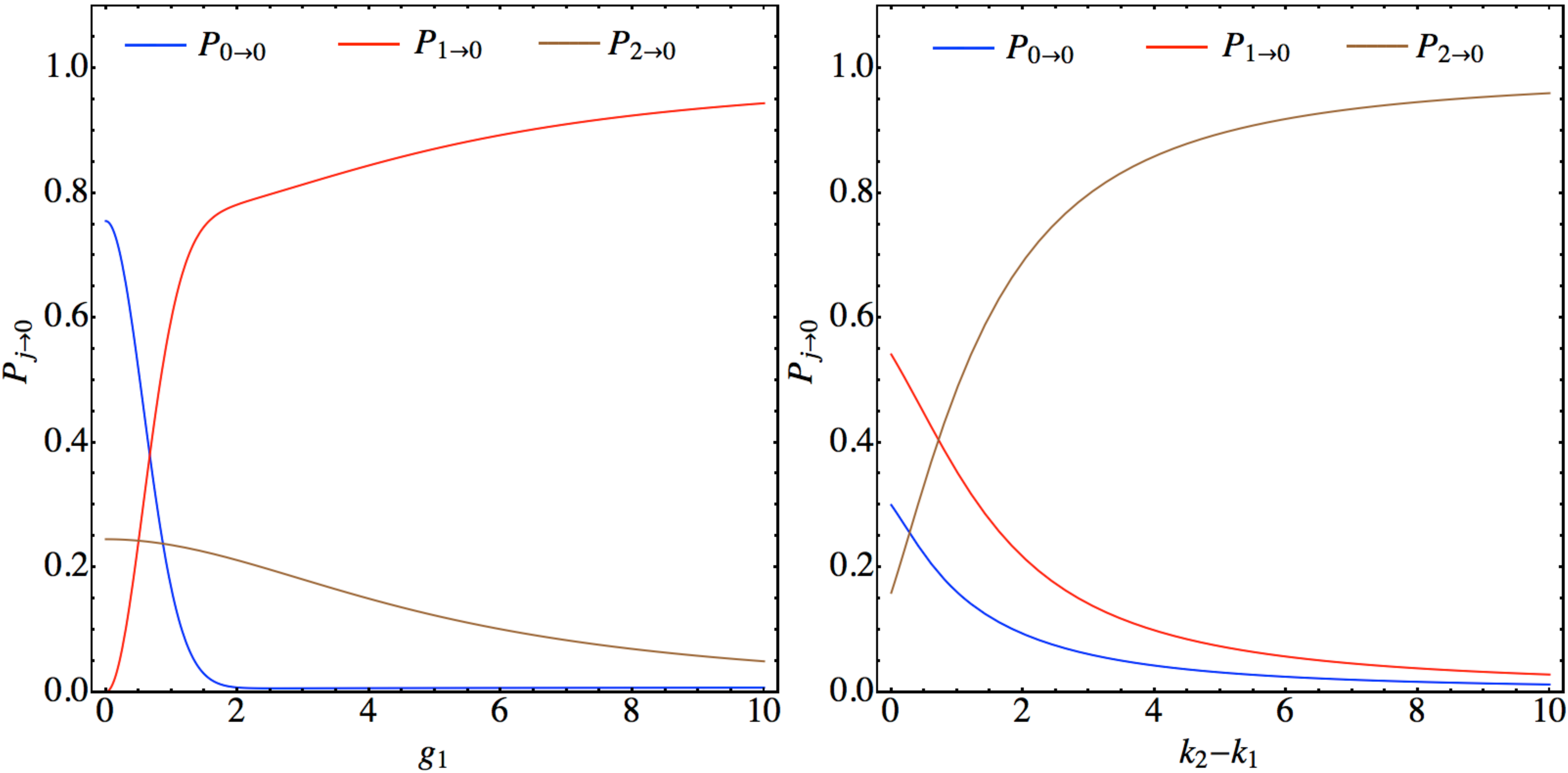}
\end{indented}
\caption{\label{figure:P1n} Transition probabilities as functions of (a)  $g_1$, and (b) the difference $k_2-k_1$, for the probabilities to find the system in the 0th level starting at arbitrary $i$th state ($i=0,1,2$). Choice of parameters:  (a) $k_1=1.57, \, k_2=12.4,\, \beta=2.02, \, g_2=0.425$;  (b) $g_1=3.4,\, g_2=1.84,\, \beta=2.02,\, \beta_2=1, \, k_1=0.27$.}
\end{figure}

Generally, the transition probabilities from any state to the $0$th state in a three-level system have the expressions:
\begin{align}
\nonumber\\
P_{00}&=\frac{1+\exp (-2\pi l_1)}{1+\exp (\pi k_1)} \frac{1+\exp (-2\pi l_2)}{1+\exp (\pi k_2)}, \label{eq:p0-3}\\
\nonumber\\
P_{10}&=\left(\frac{g_1^2}{2\beta}\right)\left(\frac{k_2/2-k_1/2}{(k_1/2+l_1)(k_1/2+l_2)}\right)\exp\left(-\pi\left(\frac{g_1^2}{\beta}+\frac{g_2^2}{\beta}\right)\right)\notag\\
&\qquad\times\frac{(\exp(\pi(k_1+2l_1))-1)(\exp(\pi(k_1+2l_2)-1))}{(\exp(-\pi k_1)-\exp(-\pi k_2))(\exp(\pi k_1)+1)}\label{eq:p10},\\
\nonumber \\
P_{20}&=\left(\frac{g_2^2}{2\beta}\right)\left(\frac{k_1/2-k_2/2}{(k_2/2+l_1)(k_2/2+l_2)}\right)\exp\left(-\pi\left(\frac{g_1^2}{\beta}+\frac{g_2^2}{\beta}\right)\right)\notag\\
&\qquad\times\frac{(\exp(\pi(k_2+2l_1))-1)(\exp(\pi(k_2+2l_2)-1))}{(\exp(-\pi k_2)-\exp(-\pi k_1))(\exp(\pi k_2)+1)}, \label{eq:p20}
\end{align}
where $l_1$ and $l_2$ are explicitly written in (\ref{roots3}). In \fref{figure:P1n} we show typical behavior of $P_{00}$, $P_{10}$, and $P_{20}$ as functions of one of the coupling constants and the difference of parameters $k_2-k_1$. In particular, \fref{figure:P1n}(b) shows that the separation between band levels, which is controlled by the value of $k_2-k_1$, considerably influences transition probabilities. This behavior is in sharp contrast with the conclusion that one can derive from the previously known Demkov-Osherov model \cite{do}. 

\section{Conclusion}

We identified and solved the model of nonadiabatic transitions in an explicitly driven multistate Landau-Zener-Coulomb-like system. Our model describes transitions between a time-dependent energy level that interacts with a band of states whose adiabatic energies change non-linearly with time. We showed that a nonlinear level crossing results in saturation of the survival probability at large coupling constants. This behavior is similar to the one found in the model in \cite{sinitsyn-13prl1}. However, unlike \cite{sinitsyn-13prl1}, our model shows that the decay of the survival probability in the large-$N$ limit generally happens if the separation between avoided crossing points becomes substantial.

Finally, we note that many properties of our model remain unstudied. For example, it should be interesting to consider a continuous limit of simultaneously large $N$ and low  $|g_i^2/\beta|$ values. This limit has been encountered in several applications \cite{lev}. Moreover, the case of all negative parameters $k_i$ corresponds to the adiabatic transitions between a single level and a band without going through any avoided level crossings, as shown in \fref{rays5}. Such a regime can be found in many previously studied mesoscopic systems \cite{lev} with nonadiabatic transitions. 

\appendix
\section{Calculations}
In the this appendix, we introduce the notation and theory used in \cite{luke} to describe the Meijer's G functions, and then we show the calculations referenced in the text.

\subsection{Shorthand}
\subsubsection*{Preliminary shorthand}
We follow the conventions of \cite{luke}, where we notate
\begin{equation}
\MeijerG*{m}{n}{p}{q}{a_p}{b_q}{z}:=\MeijerG*{m}{n}{p}{q}{a_1,\ldots,a_p}{b_1,\ldots,b_q}{z},
\end{equation}
and in general, when one expects a list of $p$ arguments, $a_p+\sigma$ will signify $\{a_1+\sigma,\ldots,a_p+\sigma\}$, for some constant $\sigma$.

We will use the following notation for sets as arguments in hypergeometric functions that follow: for a fixed $h$, we define $(1-a_h+a_p)^*=\{1-a_h+a_1,\ldots,1-a_h+a_{h-1},1-a_h+a_{h+1},\ldots,1-a_h+a_p\}$, and $(1-b_h+b_q)^*$ is defined similarly.
\subsubsection*{Symbols}
We define the following constants, which come from Section 5.7 of \cite{luke}, equations 1 and 13:
\begin{equation}
\eqalign{\Theta_1=\sum_{h=1}^q b_h,\quad\Lambda_1=\sum_{h=1}^p a_h,\\
\quad\sigma=q-p,\quad\nu=q-m-n,\quad\rho=m+n-\frac{1}{2}(p+q),}
\end{equation}
(for $k\geq 2$, there are constants $\Theta_k$ and $\Lambda_k$ which are defined recursively in section 5.11.5 of \cite{luke} -- we will not use these constants) and we define $\theta$ by the equation
\begin{equation}
\sigma\theta=\frac{1}{2}(1-\sigma)+\Theta_1-\Lambda_1.
\end{equation}

Finally, we define the following two constants (from Section 5.9 of \cite{luke}, equations 1 and 2):
\begin{equation}
\Delta(t)=(-1)^{\nu+1}\left(\prod_{\substack{j=1\\j\neq t}}^n\Gamma(a_t-a_j)\Gamma(1+a_j-a_t)\right)\left(\prod_{j=m+1}^q\Gamma(a_t-b_j)\Gamma(1+b_j-a_t)\right)^{-1}
\end{equation}
and
\begin{equation}
A=(-1)^\nu(2\pi i)^{-\nu}\exp\left[i\pi\left(\sum_{j=1}^n a_j-\sum_{j=m+1}^q b_j\right)\right].
\end{equation}
\subsubsection*{Functions}
We first define the $H$ function from Section 5.7 of \cite{luke}, equation 13:
\begin{equation}
H_{p,q}(z)=\frac{(2\pi)^{(\sigma-1)/2}}{\sigma^{1/2}}\exp\left(-\sigma(z)^{1/\sigma}\right)z^{\theta}\sum_{k=0}^\infty M_kz^{-k/\sigma},
\end{equation}
where $M_0=1$ and for $k\geq 1$, $M_k$ is independent of $z$ and defined in terms of $\Theta_k$, $\Lambda_k$, and $\sigma$.

Next we define the $E$ function from Section 5.7 of \cite{luke}, equation 7:
\begin{equation}
E_{p,q}(a_t;z)=\frac{z^{a_t-1}\prod_{j=1}^q\Gamma(1+b_j-a_t)}{\prod_{j=1}^p\Gamma(1+a_j-a_t)}{}_qF_{p-1}(1+b_q-a_t;(1+a_p-a_t)^*;-1/z).
\end{equation}

\subsection{Asymptotics and Identities}
In Section 5.10 of \cite{luke}, equation 12, we have the following divergent asymptotic for $|z|\rightarrow\infty$, $\arg z=\rho\pi$:
\begin{equation}\label{asymp1}
\MeijerG*{m}{n}{p}{q}{a_p}{b_q}{z}\sim AH_{p,q}(ze^{i\pi\nu})+\sum_{t=1}^n\exp(-i\pi a_t(\nu+1))\Delta(t)E_{p,q}(a_t;ze^{i\pi(\nu+1)}).
\end{equation}

We also mention that from the series definition of ${}_pF_q$ given in Section 3.2 of \cite{luke}, equation 2, we know that ${}_pF_q\rightarrow 1$ as $z\rightarrow 0$.

We use the following identities:

In Section 5.4 of \cite{luke}, equation 13, we have
\begin{equation}\label{deriv1}
\frac{d}{dz}\left[z^{-b_h}\MeijerG*{m}{n}{p}{q}{a_p}{b_q}{z}\right]=z^{-1-b_h}\MeijerG*{m}{n}{p}{q}{a_p}{b_1,\ldots,b_{h-1},b_{h+1},\ldots,b_{q},1+b_h}{z},
\end{equation}
for $m<h$.

In Section 5.4 of \cite{luke}, equation 12, we have
\begin{equation}\label{deriv1a}
\frac{d}{dz}\left[z^{-b_1}\MeijerG*{m}{n}{p}{q}{a_p}{b_q}{z}\right]=-z^{-1-b_1}\MeijerG*{m}{n}{p}{q}{a_p}{1+b_1,b_2,\ldots,b_q}{z}.
\end{equation}

In Section 5.4 of \cite{luke}, equation 17, we have
\begin{equation}\label{deriv2}
z^k\frac{d^k}{dz^k}\left[\MeijerG*{m}{n}{p}{q}{a_p}{b_q}{z}\right]=\MeijerG*{m}{n+1}{p+1}{q+1}{0,a_p}{b_q,k}{z}.
\end{equation}

In Section 3.4 of \cite{luke}, equation 1, we have\footnote{We note that Luke denotes $(x)_n=x(x+1)\cdots(x+n-1)$ as the rising factorial, and in order to be consistent with calculations in the main text and standard convention in combinatorial mathematics we have chosen instead to denote $(y)_n=y(y-1)\cdots(y-n+1)$ as the falling factorial. We note the obvious $(y)_{n,\textrm{falling}}=(y-n+1)_{n,\textrm{rising}}$.}
\begin{equation}\label{deriv3}
\frac{d^k}{dz^k}\left[{}_pF_q(a_p;b_q;z)\right]=\frac{(a_p+k-1)_k}{(b_q+k-1)_k}{}_pF_q(a_p+k;b_q+k;z),
\end{equation}
where $(x)_k=x(x-1)\cdots(x-k+1)$ denotes the falling factorial.

The last identity we state here is in Section 5.2 of \cite{luke}, equation 12, which gives a relation between certain parameters of the Meijer's G function and the generalized hypergeometric function:
\begin{equation}\label{elem3}
\MeijerG*{1}{n}{p}{q}{a_p}{b_q}{z}=\frac{\prod_{j=1}^n\Gamma(1+b_1-a_j)z^{b_1}}{\prod_{j=2}^q\Gamma(1+b_1-b_j)\prod_{j=n+1}^p\Gamma(a_j-b_1)}{}_pF_{q-1}(1+b_1-a_p;(1+b_1-b_q)^*;(-1)^{p-1-n}z),
\end{equation}
for $p<q$ (or $p=q$ and $|z|<1$, but this case does not concern us).

\subsection{Calculations}\label{calcs}
\begin{enumerate}
\item\label{calc1} The claimed behavior of the solutions in \eref{eq:b0fundset} as $t\rightarrow 0$ follows directly from the relation \eref{elem3} and the fact that ${}_NF_N\rightarrow 1$ as $t\rightarrow 0$.
\item\label{calc2} From the result \eref{calc1}, we know that the solution $b_0$ in \eref{eq:b0firstsol} has the property $b_0(t\rightarrow 0)<\infty$. Combining the the fact that the $k$th derivative of a hypergeometric function is a multiple of another hypergeometric function, from \eref{deriv3}, and the fact that ${}_NF_N\rightarrow 1$ as $t\rightarrow 0$, we have that
\begin{equation}
\left.\frac{d^k}{dt^k}b_0\right|_{t\rightarrow 0}<\infty.
\end{equation}
\item\label{calc3} Here we have initially populated $b_0$ with probability 1. We are interested in the asymptotics of $\MeijerG*{1}{N}{N}{N+1}{\vec{\xi_N}}{0,\vec{h_N}}{\beta it}$. Here we have $\sigma=1$, $\nu=0$, and $\rho=1/2$. We also have $\Theta_1=\sum h_j$ and $\Lambda_1=\sum\xi_j$. Hence $\theta=\Theta_1-\Lambda_1=-i\sum g_j^2/2\beta$. We have the constant
\begin{equation}
A=\exp\left[i\pi\left(\Lambda_1-\Theta_1\right)\right]=\exp\left(-\pi\sum_{j=1}^N\frac{g_j^2}{2\beta}\right),
\end{equation}
and the $H$ function becomes
\begin{equation}
H_{p,q}(z)=e^{-z}z^{-i\sum g_j^2/2\beta}\sum_{k=0}^\infty M_kz^{-k}.
\end{equation}
Truncating the divergent series in \eref{asymp1} to terms which do not vanish at $t\rightarrow\infty$ gives its behavior at $t\rightarrow\infty$; all terms $E_{p,q}$ drop out and the result is:
\begin{equation}
\MeijerG*{1}{N}{N}{N+1}{\vec{\xi_N}}{0,\vec{h_N}}{\beta it}\sim \exp\left(-\pi\sum_{j=1}^N\frac{g_j^2}{2\beta}\right)e^{-\beta it}\left(\beta it\right)^{-i\sum g_j^2/2\beta}.
\end{equation}
We set $i=e^{i\pi/2}$ and get
\begin{equation}
\left|\MeijerG*{1}{N}{N}{N+1}{\vec{\xi_N}}{0,\vec{h_N}}{\beta it}\right|^2\sim \exp\left(-\pi\sum_{j=1}^N\frac{g_j^2}{2\beta}\right).
\end{equation}
Using the fact that $|\Gamma(1/2+ix)|^2=\pi/\cosh(\pi x)$, we have result
\begin{equation}
|b_0(t\rightarrow\infty)|^2=\exp\left(-\pi\sum_{j=1}^N\frac{g_j^2}{2\beta}\right)\prod_{j=1}^N\left(\frac{\cosh(\pi l_j)}{\cosh(\pi k_j/2)}\right)=\prod_{j=1}^N\left(\frac{\exp(-2\pi l_j)+1}{\exp(\pi k_j)+1}\right).
\end{equation}
\item\label{calc4} Here we have again initially populated $b_0$ with probability 1. We are interested in the asymptotics of
\begin{equation}
t^{1-h_j}\MeijerG*{1}{N}{N}{N+1}{\vec{\xi_N}}{0,\vec{h_N}^j,h_j-1}{\beta it}.
\end{equation}
Here we have $\sigma=1$, $\nu=0$, and $\rho=1/2$. We also have
\begin{equation}
\Theta_1=-1+\sum_{r=1}^N h_r,\quad\Lambda_1=\sum_{r=1}^N\xi_r,\quad\theta=\Theta_1-\Lambda_1=1-i\sum_{r=1}^N\frac{g_r^2}{2\beta}.
\end{equation}
We have the constant
\begin{equation}
\Delta(s)=(-1)\left[\frac{\prod_{r=1,r\neq s}^N\Gamma(\xi_s-\xi_r)\Gamma(1+\xi_r-\xi_s)}{\prod_{r=1,r\neq j}^N\Gamma(\xi_s-h_r)\Gamma(1+h_r-\xi_s)}\right]\left[\Gamma(1+\xi_s-h_j)\Gamma(h_j-\xi_s)\right]^{-1},
\end{equation}
and the $E$ function is 
\begin{equation}
E_{N,N+1}(a_s;-\beta it)=(-\beta it)^{-\frac{1}{2}+il_s}\frac{\Gamma(1-\xi_s)\Gamma(h_j-\xi_s)\prod_{r=1}^N\Gamma(1+h_r-\xi_s)}{\prod_{r=1}^N\Gamma(1+\xi_r-\xi_s)}.
\end{equation}
If we multiply the series in \eref{asymp1} through by $t^{1-h_j}$ and then truncate the resulting series to terms which do not vanish at $t\rightarrow\infty$, we see that the $H$ function terms drop out and we are left with
\begin{equation}
\eqalign{t^{1-h_j}\MeijerG*{1}{N}{N}{N+1}{\vec{\xi_N}}{0,\vec{h_N}^j,h_j-1}{\beta it}\\
\sim-\sum_{s=1}^Ne^{-i\pi/2}e^{\pi l_s}\frac{\Gamma(1-\xi_s)}{\Gamma(1+\xi_s-h_j)}t^{1-h_j}(-\beta it)^{-\frac{1}{2}+il_s}\frac{\prod_{r=1,r\neq s}^N\Gamma(\xi_s-\xi_r)}{\prod_{r=1,r\neq j}^N\Gamma(\xi_s-h_r)}.}
\end{equation}
This gives the result
\begin{equation}
|a_j|^2\sim\frac{|Q|^2g_j^2}{2\beta}\left|\sum_{s=1}^N e^{-\pi l_s/2}(\beta t)^{il_s}\frac{\Gamma(1-\xi_s)}{\Gamma(1+\xi_s-h_j)}\left(\frac{\prod_{r=1,r\neq s}^N\Gamma(\xi_s-\xi_r)}{\prod_{r=1,r\neq j}^N\Gamma(\xi_s-h_r)}\right)\right|^2.
\end{equation}
\item\label{calc5} Here we have initially populated $a_q$ with probability 1. We are interested in the asymptotics of
\begin{equation}
\MeijerG*{1}{N}{N}{N+1}{\vec{\xi_N}}{h_r,0,\vec{h_N}^r}{\beta it}.
\end{equation}
Here we have $\sigma=1$, $\nu=0$, and $\rho=1/2$. We also have
\begin{equation}
\Theta_1=\sum_{j=1}^N h_j,\quad\Lambda_1=\sum_{j=1}^N\xi_j,\quad\theta=-i\sum_{j=1}^N\frac{g_j^2}{2\beta}.
\end{equation}
We have the constant
\begin{equation}
A=\exp\left[i\pi\left(h_r+\sum_{j=1}^N\xi_j-\sum_{j=1}^N h_j\right)\right]=e^{i\pi/2}\exp\left[\pi\left(\frac{k_r}{2}-\sum_{j=1}^N\frac{g_j^2}{2\beta}\right)\right],
\end{equation}
and the $H$ function is
\begin{equation}
H_{p,q}(z)=e^{-z}z^{-i\sum g_j^2/2\beta}\sum_{k=0}^\infty M_kz^{-k}.
\end{equation}
Then truncating \eref{asymp1} to terms which do not vanish at $t\rightarrow\infty$ gives the desired asymptotic; all $E_{p,q}$ drop out and the result is
\begin{equation}
\MeijerG*{1}{N}{N}{N+1}{\vec{\xi_N}}{h_r,0,\vec{h_N}^r}{\beta it}\sim e^{i\pi/2}e^{-\beta it}(\beta t)^{-i\sum g_j^2/2\beta}e^{-\pi\sum g_j^2/4\beta}e^{\pi k_r/2},
\end{equation}
which gives
\begin{equation}
|b_0|^2\sim\exp\left(-\pi\sum_{j=1}^N\frac{g_j^2}{2\beta}\right)\left|\sum_{r=1}^N c_{r,q}t^{ik_r/2}e^{\pi k_r/2}\right|^2.
\end{equation}

\end{enumerate}

\section{Amplitudes $a_j$ with $a_q$ initially populated}\label{amplitudeqj}

Taking $b_0$ as in \eref{eq:bqinitpop}, solving \eref{eq:elim2} for $b_j$ and then multiplying by $\tau=t^{1/2}\sqrt{2}$, we have the result
\begin{equation}
a_j=\frac{-i g_j t^{-ik_j/2}}{\sqrt{2}}\sum_{r=1}^N c_{r,q}t^{i k_r/2}\int_0^t x^{(1-h_j)-1}\MeijerG*{1}{N}{N}{N+1}{\vec{\xi_N}}{h_r,0,\vec{h_N}^r}{\beta it}.
\end{equation}
Using \eref{deriv1} and \eref{deriv1a}, we find
\begin{equation}
\eqalign{a_j=&\frac{-i g_j t^{-ik_j/2}}{\sqrt{2}}t^{1-h_j}\\
&\times\left(-c_{j,q}t^{i k_j/2}\MeijerG*{1}{N}{N}{N+1}{\vec{\xi_N}}{h_j-1,0,\vec{h_N}^j}{\beta it}+\sum_{r=1,r\neq j}^N c_{r,q}t^{i k_r/2}\MeijerG*{1}{N}{N}{N+1}{\vec{\xi_N}}{h_r,0,\vec{h_N}^{r,j},h_j-1}{\beta it}\right)}
\end{equation}

\end{document}